\documentclass{article}

\title{A Unified Understanding of Spin and Orbital Angular Momentum in the Complex Plane}
\author{Robert Ducharme}

\begin{document}

\maketitle

\centerline{2112 Oakmeadow Pl., Bedford, TX 76021}
\centerline{E-mail: robertjducharme66@gmail.com}

\begin{abstract} 
The quantum mechanical operator for angular momentum is transformed from the real plane into the complex plane. In doing so, the Cauchy-Riemann (C-R) equations are interpreted as constraint conditions defining two distinct domains where complex differentiation is permitted. It is shown each of these domains contains an orbital angular momentum contribution plus an non-orbital term that cancels out between them. It is further shown the field equations for spinning quantum particles include C-R equations that restrict the particles to a single complex constraint space. It is therefore proposed the non-orbital term in the constraint space angular momentum is the source of the spin.
\end{abstract}

\section{Introduction}
Newman \cite{ETN} has suggested spin angular momentum has a geometrical interpretation as a projection of dynamical properties of quantum particles from complex space into real space. The purpose of this paper is to further develop this idea using a direct transformation of the quantum mechanical operator $\hat{L}_3$ for measurable angular momentum from the real plane $\mathbf{R}^2$ into the complex plane $\mathbf{C}$. 

The Cauchy-Riemann (C-R) equations for $\mathbf{C}$ will be interpreted as constraint conditions defining two constraint spaces $\mathbf{C}_{+}$ and $\mathbf{C}_{-}$ where complex differentiation is permitted. Real constraint spaces have been extensively studied in constraint mechanics \cite{AK, RST, CA, WC}. The approach here is the same except the constraint spaces are complex. 

Let $(x,y)$ denote the coordinates of any point on $\mathbf{R}^2$. It is shown in section 2 that any function $\Psi(x,y)$ can be re-expressed in the form $\Psi(z,z^*)$ where $z=x_1+\imath x_2$ and $z^*=x_1-\imath x_2$. It is further shown any derivative of $\Psi$ that can be evaluated on $\mathbf{R}^2$ can be equivalently be evaluated in the $\mathbf{C}_{+}$ and $\mathbf{C}_{-}$ constraints spaces. 

In section 3 it is shown the $\hat{L}_3$ operator decomposes into $\hat{L}_{3+}$ and $\hat{L}_{3-}$ components for the angular momentum in each of the $\mathbf{C}_{+}$ and $\mathbf{C}_{-}$ constraint spaces respectively. It is also shown the $\hat{L}_{3+}$ and $\hat{L}_{3-}$ operators contain an equal and opposite term that cancels out between them even if there is no orbital angular momentum in $\mathbf{R}^2$. This indicates that fields in $\mathbf{C}_{+}$ and $\mathbf{C}_{-}$ domains can generate angular momentum through a mechanism that has no analogue in the real plane. It is inferred from this that if a particle were to be confined to either $\mathbf{C}_{+}$ or $\mathbf{C}_{-}$ constraint spaces the non-orbital term that ordinarily cancels between both constraint spaces would be exposed for measurement in experiment.  

It has just been argued that confinement of quantum particles to $\mathbf{C}_{+}$ or $\mathbf{C}_{-}$ constraint spaces is a potential source of angular momentum different in origin from orbital angular momentum. It is of interest therefore to consider the idea this unidentified form of angular momentum is spin angular momentum for two reasons. First, the generating mechanism is different from orbital angular momentum. Second, the two constraint spaces are suggestive of the two possible states of spin.

The method for connecting $\mathbf{C}_{+}$ and $\mathbf{C}_{-}$ to spinning quantum particles is to seek the signature C-R equations of these constraint spaces in the component form of the field equations that describe the particles. To be meaningful the complex constraint space for the field must determine the sign of the spin. Electromagnetic and Dirac fields are both treated in section 4. For electromagnetic waves, the C-R equations are shown to emerge in the form of the  Lorentz gauge condition for the left and right handed circular polarization states.  In this case of the Dirac field, the C-R equations are found in the component form of the Dirac equation. It is further found that spin-up and spin-down particles are confined to different constraint spaces.

\section{Transformation into Complex Coordinates}
The task ahead is to present a transformation for relating each point $(x_1, x_2)$ on a plane in real space to corresponding points $(z,z^*)$ in the complex plane. This transformation may be expressed as
\begin{equation} \label{eq: zdef}
z = x_1 + \imath x_2
\end{equation}
\begin{equation} \label{eq: zdef_conj}
z^* = x_1 - \imath x_2
\end{equation}
where $z^*$ is the complex conjugate of $z$ .

The inverses of eqs. (\ref{eq: zdef}) and (\ref{eq: zdef_conj}) can be written
\begin{equation} \label{eq: inv_x1}
x_1 = \frac{z + z^*}{2}, 
\end{equation}
\begin{equation} \label{eq: inv_x2}
x_2 = \frac{z - z^*}{2\imath}
\end{equation}
Eqs. (\ref{eq: inv_x1}) and (\ref{eq: inv_x2}) can be used alongside the chain rule for partial differentiation to derive the Wurtinger derivatives \cite{GR}:
\begin{equation} \label{eq: zdiff}
\frac{\partial}{\partial z}  
= \frac{1}{2} \left( \frac{\partial}{\partial x_1}
- \imath \frac{\partial}{\partial x_2} \right),
\end{equation}
\begin{equation} \label{eq: zdiff_conj}
\frac{\partial}{\partial z^*}  
= \frac{1}{2} \left( \frac{\partial}{\partial x_1}
+ \imath \frac{\partial}{\partial x_2} \right),
\end{equation}
Inserting eqs. (\ref{eq: zdef}) and (\ref{eq: zdef_conj}) into eqs. (\ref{eq: zdiff}) and (\ref{eq: zdiff_conj}) gives
\begin{equation} \label{eq: independence}
\frac{\partial z}{\partial z^*}=\frac{\partial z^*}{\partial z} = 0
\end{equation}
showing $z$ and $z^*$ can be varied independently of each other. 

Eqs. (\ref{eq: zdiff}) and (\ref{eq: zdiff_conj}) are only meaningful in domains where the C-R equations are satisfied, otherwise, the result of the differentiation will depend on the direction it is performed. The C-R equations needed to ensure complex differentiation with respect to $z$ is meaningful take the form
\begin{equation} \label{eq: cr}
\frac{\partial \Psi}{\partial x_1} + \imath \frac{\partial \Psi}{\partial x_2} = 2\frac{\partial \Psi}{\partial z^*}  = 0, 
\end{equation}
It is clear that a function $\Psi(z,z^*)$ can only satisfy eqs. (\ref{eq: cr}) if $z^*$ is held constant. This implies $\frac{\partial}{\partial z}$ cannot exist except in the constraint space 
\begin{equation} \label{eq: dom1}
\mathbf{C}_{+} = \{ z, z^* \in \mathbf{C} | z^* = \xi^* \}
\end{equation}
where $\xi = \xi_a + \imath \xi_b$ is an arbitrary complex constant and $\xi^*$ is the complex conjugate of it.

The form of the C-R equations needed to ensure complex differentiation with respect to $z^*$ is meaningful is simply the complex conjugate of eqs. (\ref{eq: cr}). This can be written
\begin{equation} \label{eq: cr_conj}
\frac{\partial \Psi}{\partial x_1} - \imath \frac{\partial \Psi}{\partial x_2} = 2\frac{\partial \Psi}{\partial z}  = 0, 
\end{equation}
It is clear that a function $\Psi(z,z^*)$ can only satisfy eqs. (\ref{eq: cr_conj}) if $z$ is held constant. This implies $\frac{\partial}{\partial z^*}$ cannot exist except in the constraint space 
\begin{equation} \label{eq: dom2}
\mathbf{C}_{-} = \{ z, z^* \in \mathbf{C} | z = \xi \}
\end{equation}
It has been shown overall that the C-R eqs. (\ref{eq: cr}) and (\ref{eq: cr_conj}) can be interpreted as constraint conditions defining the $\mathbf{C}_{+}$ and $\mathbf{C}_{-}$ constraint spaces where any analytic function can be differentiated with respect to $z$ and $z^*$ respectively.

The foregoing argument implies any derivative $\frac{\partial \Psi}{ \partial x_i}$ that can be evaluated in $\mathbf{R}^2$ can be equivalently evaluated in the $\mathbf{C}_{+}$ and $\mathbf{C}_{-}$ constraint spaces. The method is in three parts. First, rewrite $\Psi(x_1,x_2)$ in the form  $\Psi(z,z^*)$ using eqs. (\ref{eq: inv_x1}) and (\ref{eq: inv_x2}). Second, evaluate $\frac{\partial \Psi}{ \partial z}$ and $\frac{\partial \Psi}{ \partial z^*}$ using the normal rules of differentiation taking into account eq. (\ref{eq: independence}). Thirdly, calculate the two components of $\frac{\partial \Psi}{\partial x_i}$ from the inverse form of eqs. (\ref{eq: zdiff}) and (\ref{eq: zdiff_conj}):
\begin{equation} \label{eq: invzdiff}
\frac{\partial}{\partial x_1}  
=  \frac{\partial}{\partial z}
+ \frac{\partial}{\partial z^*},
\end{equation}
\begin{equation} \label{eq: invComplexDiff2}
\frac{1}{\imath}\frac{\partial}{\partial x_2}  
=  \frac{\partial}{\partial z}
- \frac{\partial}{\partial z^*},
\end{equation}
The method is, of course, readily validated using any explicit form for $\Psi(x_1,x_2)$ since differentiation must give the same result irrespective of whether it is performed in $\mathbf{R}^2$ or in $\mathbf{C}_{+}$ and $\mathbf{C}_{-}$.

Derivatives of any function $\Psi$ confined to either of the $\mathbf{C}_{+1}$ or $\mathbf{C}_{-1}$ constraint spaces can be evaluated using eqs. (\ref{eq: invzdiff}) and (\ref{eq: invComplexDiff2}) to give
\begin{equation} \label{eq: constrained_deriv1}
\left(\frac{\partial^{m+n} \Psi}{\partial x_1^m \partial x_2^n} \right)_{+} =  (+\imath)^n \left(\frac{\partial}{\partial z} \right)^{m+n} \Psi, 
\end{equation}
\begin{equation} \label{eq: constrained_deriv2}
\left(\frac{\partial^{m+n} \Psi}{\partial x_1^m \partial x_2^n} \right)_{-} =  (-\imath)^n \left(\frac{\partial}{\partial z^*} \right)^{m+n} \Psi, 
\end{equation}
where $m, n = 0, 1, 2 ...$ and the $\pm$ subscripts indicate the constraint space confining the operation. It is also useful to transform both of these results back into real space using eqs. (\ref{eq: zdiff}) leading to
\begin{equation} \label{eq: constrained_deriv3}
\left(\frac{\partial^{m+n} \Psi}{\partial x_1^m \partial x_2^n} \right)_{+} =  (+\imath)^n  \left[\frac{1}{2}\left(\frac{\partial}{\partial x_1} - \imath \frac{\partial}{\partial x_2} \right) \right]^{m+n} \Psi, 
\end{equation}
\begin{equation} \label{eq: constrained_deriv4}
\left(\frac{\partial^{m+n} \Psi}{\partial x_1^m \partial x_2^n} \right)_{-} =  (-\imath)^n \left[\frac{1}{2}\left(\frac{\partial}{\partial x_1} + \imath \frac{\partial}{\partial x_2} \right) \right]^{m+n} \Psi
\end{equation}
Eqs. (\ref{eq: constrained_deriv3}) and (\ref{eq: constrained_deriv4}) show that any function $\Psi(x_1,x_2)$ will satisfy Laplace's equation
\begin{equation} \label{eq: laplace}
\left(\frac{\partial^2 \Psi}{\partial x_1^2} + \frac{\partial^2 \Psi}{\partial x_2^2} \right)_{\pm} =  0
\end{equation}
in both the $\mathbf{C}_{+}$ and $\mathbf{C}_{-}$ constraint spaces irrespective of the form of $\Psi(x_1,x_2)$. This is, in fact, just an alternative statement of the C-R equations.

\section{Angular Momentum in the Complex Plane}
It is instructive to investigate the transformation of the quantum mechanical operator for observable angular momentum \cite{DFL} from the real plane into the $\mathbf{C}_{+}$ and $\mathbf{C}_{-}$ constraint spaces. This operator takes the form
\begin{equation} \label{eq: ang_mom_opr}
\hat{L}_3(x_1,x_2) =  \frac{\hbar}{\imath}\left( x_1 \frac{\partial}{\partial x_2} - x_2 \frac{\partial}{\partial x_1} \right)
\end{equation}
where $\hbar$ is Planck's constant divided by $2 \pi$.

Inserting eqs. (\ref{eq: inv_x1}), (\ref{eq: inv_x2}), (\ref{eq: invzdiff}) and (\ref{eq: invComplexDiff2}) into eq. (\ref{eq: ang_mom_opr}) gives
\begin{equation} \label{eq: ang_mom_planes}
\hat{L}_3 = \hat{L}_{3+} + \hat{L}_{3-}
\end{equation}
where
\begin{equation} \label{eq: ang_mom_z_1}
\hat{L}_{3+} = \hbar z \frac{\partial}{\partial z}
\end{equation}
\begin{equation} \label{eq: ang_mom_z_2}
\hat{L}_{3-} = -\hbar z^* \frac{\partial}{\partial z^*}
\end{equation}
The understanding here is that the $\hat{L}_{3+}$ and $\hat{L}_{3-}$ operators give the angular momentum in the $\mathbf{C}_{+}$ and $\mathbf{C}_{-}$ constraint spaces respectively. Eq. (\ref{eq: ang_mom_planes}) shows the total angular momentum in the real plane is equal to the sum of contributions from the two complex constraint spaces. Eqs. (\ref{eq: ang_mom_z_1}) and (\ref{eq: ang_mom_z_2}) are readily transformed into real coordinates using eqs. (\ref{eq: zdef}), (\ref{eq: zdef_conj}), (\ref{eq: zdiff}) and (\ref{eq: zdiff_conj}) to give
\begin{equation} \label{eq: ang_mom_z_3}
\hat{L}_{3+} = \frac{1}{2}(\hat{L}+\hat{S})\Psi
\end{equation}
\begin{equation} \label{eq: ang_mom_z_4}
\hat{L}_{3-} = \frac{1}{2}(\hat{L}-\hat{S})\Psi
\end{equation}
where
\begin{equation} \label{eq: ang_mom_spin}
\hat{S} = \hbar \left(x_1 \frac{\partial}{\partial x_1} + x_2 \frac{\partial}{\partial x_2} \right)
\end{equation}
is an operator for a form of angular momentum that has a different origin from orbital angular momentum. It is of interest next to consider the idea that the $\hat{S}$ operator is related to spin angular momentum. The reasoning is that although the $\hat{S}$ term cancels out between the $\mathbf{C}_{+}$ and $\mathbf{C}_{-}$ constraint spaces, this hidden term might be observed if a field were to be confined to just one of the two constraint spaces. In seeking the connection to spin the logical approach is therefore to determine if spinning particles satisfy C-R equations for either $\mathbf{C}_{+}$ or $\mathbf{C}_{-}$ domains such that the constraint space encapsulating the field determines the sign of the spin for all the quantum particles it contains.

\section{Identifying the Complex Constraint Spaces of Spinning Particles}
The objective of this section is to show the field equations for electromagnetic radiation and Dirac particles naturally include the C-R equations for the $\mathbf{C}_{+}$ and $\mathbf{C}_{-}$ constraint spaces. It is also to show the constraint space of the field reflects the sign of the spin of the particles in it.

Electromagnetic radiation \cite{IZ} can be represented using a 4-potential $A_\mu (x_\nu)$ where $\mu, \nu = 0,1,2,3$ and $x_\mu $ is position in Minkowski 4-space. The classical field equations for $A_\mu$  consist of Maxwell's equations 
\begin{equation} \label{eq: maxwell} 
\frac{\partial^2 A_\mu}{\partial x_1^2} + \frac{\partial^2 A_\mu}{\partial x_2^2} + \frac{\partial^2 A_\mu}{\partial x_3^2} - \frac{1}{c^2}\frac{\partial^2 A_\mu}{\partial t^2} = 0
\end{equation}
(having put $t=x_0$) and the Lorenz gauge condition
\begin{equation} \label{eq: lorenz_gauge} 
\frac{\partial A_1}{\partial x_1} + \frac{\partial A_2}{\partial x_2} + \frac{\partial A_3}{\partial x_3} - \frac{\partial A_0}{\partial t} = 0
\end{equation}
where $c$ is the velocity of light. 

The solutions to Maxwell's equations (\ref{eq: maxwell}) for plane electromagnetic waves moving in the $x_3$-direction can be written in the general form
\begin{equation} \label{eq: em_node} 
A_{\mu}^{\beta} = \epsilon_{\mu}^{\beta} \psi_E(x_\nu)
\end{equation}
where $\epsilon_{\mu}^\beta$ is the polarization vector for polarization state $\beta$ and $\psi_E(x_\nu)$ is a function of space and time. It will be convenient to set
\begin{equation} \label{eq: right_circular} 
\epsilon_{\mu}^{\beta} = \frac{1}{\sqrt{2}}
\left( \begin{array}{c}
1 \\
\beta \imath \\
0 \\
0
\end{array} \right)
\end{equation}
denoting the left $\beta = 1$ and right $\beta = -1$ circular polarization states.

Inserting eq. (\ref{eq: em_node}) into (\ref{eq: lorenz_gauge}) gives
\begin{equation} \label{eq: cr_em}
\frac{\partial \psi_E}{\partial x_1} + \imath \frac{\partial \psi_E}{\partial x_2}  = 0, 
\end{equation}
for left circularly polarized light; and
\begin{equation} \label{eq: cr_conj_em}
\frac{\partial \psi_E}{\partial x_1} - \imath \frac{\partial \psi_E}{\partial x_2}  = 0, 
\end{equation}
for right circularly polarized light. Eqs. (\ref{eq: cr_em}) and (\ref{eq: cr_conj_em}) are the C-R equations for the $\mathbf{C_{+}}$ and $\mathbf{C_{-}}$ constraint spaces. It is known left and right circular polarized photons have a respective spin angular momentum of $+\hbar$ and $-\hbar$. It is inferred on this basis that positive spin photons only occupy the $\mathbf{C_{+}}$ constraint space and negative spin photons are restricted to the $\mathbf{C_{-}}$ constraint space.

The free-field equation for Dirac particles is
\begin{equation} \label{eq: dirac} 
\gamma_1 \frac{\partial \Psi}{\partial x_1} + \gamma_2 \frac{\partial \Psi}{\partial x_2} + \gamma_3 \frac{\partial \Psi}{\partial x_3} - \gamma_0\frac{1}{c} \frac{\partial \Psi}{\partial t} = \frac{\imath m_0c^2}{\hbar}\Psi
\end{equation}
where $\gamma_{\mu}$ are Dirac matrices and $m_0$ is the rest mass of each particle. The positive-energy solutions to this equation for particles moving in the $x_3$-direction are known to take the general form
\begin{equation} \label{eq: dirac_node} 
\Psi_{\beta} = u_{\beta} \psi_D(x_\mu)
\end{equation}
where
\begin{eqnarray} \label{eq: spinors} 
u_+=\left( \begin{array}{c}
1 \\
0 \\
0 \\
0
\end{array} \right) \quad
u_-=\left( \begin{array}{c}
0 \\
1 \\
0 \\
0
\end{array} \right) 
\end{eqnarray} 
are respective bi-spinors for spin-up and spin-down particles.

Inserting eqs. (\ref{eq: dirac_node}) into eq. (\ref{eq: dirac}) leads to the component form of the Dirac equation. These components include
\begin{equation} \label{eq: cr_dirac}
\frac{\partial \psi_D}{\partial x_1} + \imath \frac{\partial \psi_D}{\partial x_2}  = 0, 
\end{equation}
for spin-up particles; and
\begin{equation} \label{eq: cr_conj_dirac}
\frac{\partial \psi_D}{\partial x_1} - \imath \frac{\partial \psi_D}{\partial x_2}  = 0, 
\end{equation}
for spin-down particles. Eqs. (\ref{eq: cr_em}), (\ref{eq: cr_conj_em}), (\ref{eq: cr_dirac}) and (\ref{eq: cr_conj_dirac}) show collectively that the fields of both spinning photons and Dirac particles are naturally confined to either the $\mathbf{C_{+}}$ or $\mathbf{C_{-}}$ constraint space as hypothesized in section 3. It is also clear the two constraint spaces are experimentally distinguishable through the sign of the spin of the quantum particles that occupy them.

\section{Summary}
It has been shown the quantum mechanical operator for measurable angular momentum can be represented in the complex plane. In this coordinate system the measurable angular momentum is sum of components from two different constraint spaces where the C-R equations permit complex differentiation. An interesting feature of this representation is the presence of an extra term in the angular momentum operators for both constraint spaces that cancels out between them. This extra term has been proposed to be related to spin angular momentum since it is clearly generated through a different mechanism than orbital angular momentum.

In connecting complex constraint spaces to spin angular momentum, it has been shown the C-R equations that define the constraint spaces are naturally present in the field equations for electromagnetic waves (circular polarization) and Dirac particles. The reasoning here is that the spin term in the constraint space angular momentum is usually hidden because it cancels out between the two constraint spaces. The affect of the C-R equations on the fields is therefore to confine them to a single constraint space thus exposing the spin in it. It has further been shown each constraint space only contains spinning particles of one sign such that the sign of the spin angular momentum on quantum particles identifies the complex constraint space that confines them.

\end{document}